\documentclass[a4paper,10pt]{article} 

\onecolumn
\def\sloppy{\tolerance=100000\hfuzz=\maxdimen \vfuzz=\maxdimen}
\usepackage{graphicx}
\usepackage{amssymb}
\usepackage{amsmath}
\usepackage{amsfonts}
\usepackage{amsthm}
\usepackage{xfrac}
\usepackage{mathtext}
\usepackage{xcolor}

\newcommand{\rme}{\mathrm{e}}
\newcommand{\rmi}{\mathrm{i}}
\newcommand{\rmd}{\mathrm{d}}
\newcommand{\Tr}{\mathrm{Tr}}
\newcommand{\bg}{\boldsymbol}
\newcommand{\llangle}{\langle\hspace*{-0.95mm}\langle}
\newcommand{\rrangle}{\rangle\hspace*{-0.95mm}\rangle}

\vspace{-8ex}\date{}

\begin{document}
\sloppy

\title{\Large\bf Scaling behavior and phases of nonlinear sigma model\\
on real Stiefel manifolds near two dimensions} 

\author{{\small A.M. Gavrilik}\footnote{omgavr@bitp.kyiv.ua}\ \
{\small and A.V. Nazarenko}\footnote{nazarenko@bitp.kyiv.ua}\\
{\small\it Bogolyubov Institute for Theoretical Physics of NAS of Ukraine,}\\
{\small\it 14-b, Metrolohichna str., Kyiv, 03143, Ukraine}}

\maketitle

\begin{abstract}
For a quasi-two-dimensional nonlinear sigma model on the real Stiefel
manifolds with a generalized (anisotropic) metric, the equations of a two-charge
renormalization group (RG) for the homothety and anisotropy of the metric as
effective couplings are obtained in one-loop approximation. Normal coordinates
and the curvature tensor are exploited  for the renormalization of the metric.
The RG trajectories are investigated and the presence of a fixed point common
to four critical lines or four phases (tetracritical point) in the general case,
or its absence in the case of Abelian structure group, is established.
For the tetracritical point, the critical exponents are evaluated and 
compared with those known earlier for simpler particular case.

\end{abstract}

\section{Introduction}

The nonlinear sigma model (NLSM), first introduced in~\cite{GL60} for a
description of pion properties, has found gene\-ralizations which have become
a very efficient tool in exploring diverse phenomena in quantum physics,
including, e.g.,~magnetic systems~\cite{Pol75,VZB71,BZJ76} as well as the conductivity
of disordered systems~\cite{Hou80}. Soon after~\cite{GL60}, the~importance
of using (differential) geometric methods in the construction and study of
general phenomenological Lagrangians (NLSMs) was recognized
(see~\cite{CWZ69,CCWZ69,Vol73}). Moreover, the~importance of symmetric homogeneous
(or coset) spaces was emphasized in~\cite{Vol73} as well as in~\cite{Hou80,HLN80}.
On the other hand, a~few works~\cite{Gav86,Gav89,Gav00} paid some attention to
the potential role of nonsymmetric coset spaces as target spaces for the respective
nonlinear sigma~models.

Closely related to the well-known real Grassmann manifolds ${\rm Gr}_k(\mathbb{R}^N)$
are their well-known extensions~\cite{Hus}, such as the real Stiefel manifolds
$V_k(\mathbb{R}^N)$.
The latter are nothing but principal fiber bundles with a base being the Grassmannian
${\rm Gr}_k(\mathbb{R}^N)$ and the structure group $SO(k)$. We are interested in
(quasi) two-dimensional sigma models---nonlinear field models with values in the Stiefel
(or, in~particular, Grassmann) manifolds. We note that the quantum properties of
Grassmannian sigma models are quite well studied: in the two-dimensional case, such
nonlinear models are characterized by asymptotic freedom~\cite{Pol75,KS80},
and~if we extend to the dimension $d=2+\varepsilon$, the~models have a two-phase structure
with a bicritical point~\cite{Hi80}. This behavior, together with the use of the
replica method and replica limit, allowed the description (see, e.g.,~\cite{Hou80})
of~the (de)localization of electrons in disordered systems (or random potential)
\cite{And79}. As~for the classical two-dimensional Grassmannian models,
their integrability was proven for these nonlinear field models~\cite{Zah78}.

Nonlinear sigma models on real Stiefel manifolds have radically different
properties, both in the classical and especially quantum consideration~\cite{Gav89,Gav91}, 
both for $d=2$ and \mbox{$d=2+\varepsilon$.} In~particular, in~contrast to the single-charge
RG (scaling) behavior of Grassmannian  sigma models, the~class of $V_k(\mathbb{R}^N)$
sigma models is characterized by a two-charge behavior, and~the question of the presence
of asymptotic freedom in them at $d=2$ becomes quite nontrivial~\cite{Gav89,Gav91}.

The extension of the ${\rm Gr}_k(\mathbb{R}^N)$ sigma model to the $d=(2+\varepsilon)$-dimensional
nonlinear $V_k(\mathbb{R}^N)$-model (of the fields defined on real Stiefel manifolds)
is also very desirable in view of the following important reasons. {\it First},
since Grassmannians are Einstein manifolds, the~single (effective) charge or coupling
in the associated quantum sigma model is of pure geometrical origin---it is
the homothety of metrics, i.e., uniform (or isotropic) scaling of a fixed metric.
On the contrary, in~the extended models on the spaces $V_k(\mathbb{R}^N)$
the latter in general are not Einstein manifolds, and~we come inevitably to
the enriched set of couplings: besides homothety, a certain anisotropy of
metrics comes into play. Then, the~Einsteinian property may occur only very randomly
(for one or two special values of the anisotropy). {\it Second}, because~of such
nontrivial geometry of $V_k(\mathbb{R}^N)$, the~related nonlinear sigma models in
$d=2+\varepsilon$ do manifest more complicated critical behavior: besides/instead
of bicritical point, a tetracritical point may also appear. {\it The third} property is
very important from the viewpoint of possible physical applications. Namely, we have
in mind applying $V_k(\mathbb{R}^N)$ sigma models to a more complex~situation when
the system manifesting (de)localization of electrons may also participate in
superconductivity. It is clear that the involvement of an additional order parameter could
then be related to the presence of a second coupling in the Stiefel model. Finally,
the electron-hole symmetry that finds its natural reflection in the $k\leftrightarrow N-k$
symmetry of Grassmannian sigma models (see, e.g.,~\cite{Hi80}) is known to be inevitably
broken within the superconductivity context as noted, e.g.,~in~\cite{Hirsch,Ghosh}.
However, just the absence of $k\leftrightarrow N-k$ symmetry is an intrinsic feature
of the Stiefel $V_k(\mathbb{R}^N)$ sigma~models.

Because of this, and~also in connection with the recent return of interest~\cite{Hup21,Ng22}
to the geometry and other properties of Stiefel manifolds, we will focus on the
special properties of the nonlinear sigma model in $d=2+\varepsilon$
Euclidean dimensions with the action functional ${\cal A}$ and
the Lagrangian ${\cal L}$:
\begin{equation}
{\cal A}[U]=\frac{1}{2T}\,\int {\cal L}\,\rmd^d x,\qquad
{\cal L}[U]=\Tr\left[{\bg\nabla}U^\top {\bg\nabla}U
+(\lambda-1){\bg\nabla}U^\top UU^\top{\bg\nabla}U\right],
\label{Lag}
\end{equation}
where we have introduced the temperature $T$, field $U(x)$
valued in the Stiefel manifold $V_k(\mathbb{R}^N)=SO(N)/SO(N-k)$, gradient
${\bg\nabla}=(\partial/\partial x_1,...,\partial/\partial x_d)$,
and parameter (of the metric anisotropy) $\lambda>0$.

\section{Some characteristics of Stiefel~manifold}

A real Stiefel manifold $V_k(\mathbb{R}^N)$ is the set of all $k$-frames formed
by sets of orthonormal vectors in the space $\mathbb{R}^N$. Any element
$U\in V_k(\mathbb{R}^N)$, given by an $N\times k$-matrix, satisfies the orthogonality
condition $U^\top U=I_k$, where $I_k$ is the identity $k\times k$-matrix
and $\top$ denotes the transpose. On~such manifolds, the~group $SO(N)$ acts transitively
as the {\it isometry} group. The~{\it isotropy} group corresponding to the ``origin''
${\cal O}=\{{\bf e}_1,{\bf e}_2,...,{\bf e}_k\}$ consists of matrices of the
form ${\rm diag}(I_k,H)$, where $H\in SO(N-k)$. This allows us to identify
the manifold $V_k(\mathbb{R}^N)$ with the quotient space $SO(N)/SO(N-k)$
with $\dim V_k(\mathbb{R}^N)=Nk-k(k+1)/2$. Thus, $V_k(\mathbb{R}^N)$
is defined as a submanifold of the space $\mathbb{R}^{N\times k}$. In~particular,
$V_1(\mathbb{R})=\{1,-1\}$; $V_1(\mathbb{R}^N)=S^{N-1}$ is an $(N-1)$-dimensional
sphere; $V_N(\mathbb{R}^N)=O(N)$; $V_{N-1}(\mathbb{R}^N)=SO(N)$; and
$V_2(\mathbb{R}^4)$ is the first nontrivial~case.

{\bf Tangent space.} Let ${\cal TM}$ be the tangent space for ${\cal M}=V_k(\mathbb{R}^N)$.
We require that each tangent vector $\xi\in{\cal T}_U{\cal M}$ at a point $U\in{\cal M}$
belongs to the {\it horizontal} subspace, i.e.,~it has the following
structure~\cite{Hup21,Ng22,Edel98}:
\begin{eqnarray}
&\xi=U\omega_\xi+F_\xi,&
\label{xi}\\
&\omega_\xi=\frac{1}{2}(U^\top\xi-\xi^\top U)\in\frak{o}(k),\qquad
F_\xi=\Pi\xi,&
\nonumber
\end{eqnarray}
where the projector $\Pi=I_N-UU^\top$, and~$\omega^\top_\xi=-\omega_\xi$.
Since $\omega_\xi$ consists of $k(k-1)/2$ independent matrix components and
$F_\xi$ contains $(N-k)k$ components, their total sum restores
$\dim V_k(\mathbb{R}^N)$. Conversely, the~symmetric component
$(U^\top\xi+\xi^\top U)/2$ belongs to the {\it vertical} subspace.

Tangent vectors can be defined in the respective Lie algebra
$\frak{so}(N)$, the~basis of which is formed by $N(N-1)/2$
skew-symmetric $N\times N$-matrices $X_A$:
\begin{equation}\label{base}
X_A=\left\|(X_A)_{b,c}\right\|=\left\|\delta_{b,a_1}\delta_{c,a_2}-\delta_{b,a_2}\delta_{c,a_1}\right\|,
\end{equation}
where the multi-index $A=(a_1,a_2)$ consists of the two subsets such that $1\leq a_1,a_2\leq N$.
The matrix $X_A$ has only two nonzero elements, $1$ and $-1$, at~positions $(a_1,a_2)$
and $(a_2,a_1)$, so that $X_A^\top=-X_A$ and $\Tr(X^\top_A X_A)=2$. The~basis
elements of the algebra $\frak{so}(N)$ satisfy the commutation relations:
\begin{eqnarray}
&[X_A,X_B]=f^C_{A B} X_C,&
\label{son}\\
&f^C_{A B}=\delta_{a_1,b_1}\delta^{c_1}_{a_2}\delta^{c_2}_{b_2}
+\delta_{a_2,b_2}\delta^{c_1}_{a_1}\delta^{c_2}_{b_1}
-\delta_{a_2,b_1}\delta^{c_1}_{a_1}\delta^{c_2}_{b_2}
-\delta_{a_1,b_2}\delta^{c_1}_{a_2}\delta^{c_2}_{b_1}.&
\nonumber
\end{eqnarray}

In this basis, the~orthonormal metric $g(X,Y)=-{\cal B}(X,Y)/[2(N-2)]$ based on
the bi-invariant Killing form ${\cal B}(X,Y)$, when
${\cal B}(X_A,X_B )=f_{A C}^D f_{B D}^C$, coincides with the Frobenius metric,
which is natural in the description of the $\sigma$-model (\ref{Lag}):
\begin{equation}\label{mmm}
\langle X_A,X_B\rangle_F=\frac{1}{2}\Tr(X_A^\top X_B)=\delta_{AB}.
\end{equation}
Then, the norm of the vector $\xi\in{\cal TM}$ is written as
$\|\xi\|^2_F=(1/2)\Tr(\xi^\top\xi)$.

To write the $\omega$- and $F$-components of tangent vectors in
the basis introduced, we define $X_P$ and $X_Q$ with $P=(p,q)$ and
$Q=(j,p)$, where $1\leq p,q\leq k$ and $k+1\leq j\leq N$. Although~$F\in\mathbb{R}^{N\times k}$
in (\ref{xi}), it has $(N-k)k$ independent
components and is usually included in meaningful expressions as a combination
$F^\top F$.

{\bf Metric.} The covariant metric ${\rm g}(U;\lambda)$ (metric operator)
of model (\ref{Lag}) and its action on the tangent vector
$\xi\in{\cal T}_U{\cal M}$ are defined as
\begin{equation}\label{met}
{\rm g}\xi=\xi+(\lambda-1)UU^\top\xi, \qquad
{\rm g}^{-1}\xi=\xi+(\lambda^{-1}-1)UU^\top\xi,\qquad
\det{\rm g}=\lambda^k.
\end{equation}
Note that the case $\lambda=1/2$ is called {\it canonical} in the
literature~\cite{Ng22,Edel98}.

Then the inner (scalar) product of two tangent vectors $\xi,\eta\in{\cal T}_U{\cal M}$
is given as
\begin{equation}
\langle\xi,\eta\rangle_{\rm g}=\Tr\left[\xi^\top\eta+(\lambda-1)\xi^\top UU^\top\eta\right].
\end{equation}

Since $\langle\xi,\xi\rangle_{\rm g}=\Tr(\lambda\omega^\top_\xi\omega_\xi+F^\top_\xi F_\xi)$,
according to (\ref{xi}), the~Lagrangian in (\ref{Lag}) becomes
\begin{equation}\label{LwF}
{\cal L}=\Tr(\lambda{\bg\omega}^\top_u{\bg\omega}_u+{\bf F}^\top_u {\bf F}_u).
\end{equation}
Here we use the following decomposition of the tangent vector ${\bg\nabla}U$:
\begin{eqnarray}
&{\bg\nabla}U=U{\bg\omega}_u+{\bf F}_u,&
\label{DU}\\
&{\bg\omega}_u=U^\top{\bg\nabla}U,\qquad
{\bf F}_u=\Pi{\bg\nabla}U,&
\nonumber
\end{eqnarray}
because ${\bg\nabla}U^\top U+U^\top{\bg\nabla}U=0$ due to the constraint
$U^\top U=I_k$.

Expression (\ref{LwF}) indicates the presence of two $\omega$- and $F$-subsystems.
The condition $U^\top U=I_k$ can be easily satisfied by reducing the system to an
$O(k)$-model with ${\bg\omega}_u\not=0$ and ${\bf F}_u=0$.
If ${\bg\omega}_u$ is expressed through ${\bf F}_u\not=0$, then
a model on the Grassmannian 
${\rm Gr}_k(\mathbb{R}^N)=SO(N)⁄[SO(N-k)\times SO(k)]$ is obtained.
However, our focus here is on incorporating both subsystems, utilizing
the full number of degrees of freedom, $\dim V_k(\mathbb{R}^N)$.

{\bf Linear connection and normal coordinates.} Consider the auxiliary problem of
one-dimensional evolution, which is generated by the action functional:
\begin{equation}
S=\int\rmd t\left\{\Tr\left[{\dot U}^\top {\dot U}-(\lambda-1)(U^\top {\dot U})^2\right]
+\Tr\left[H(U^\top U-I_k)\right]\right\},
\end{equation}
where $H^\top=H\in\mathbb{R}^{k\times k}$ is a Lagrange multiplier; ${\dot U}=\rmd U/\rmd t$.

By varying $U(t)$ and $H$ and using certain identities, one can obtain
the multiplier $H=-{\dot U}^\top {\dot U}+2(\lambda-1)(U^\top {\dot U})^2$ and
the equation of the geodesic~\cite{Hup21}:
\begin{equation}
{\ddot U}+\Gamma({\dot U},{\dot U})=0.
\label{geo1}
\end{equation}
To write the latter, we used the expression for the Christoffel function
(symbol of the second kind), which defines the linear connection at the point $U\in{\cal M}$:
\begin{equation}
\Gamma(\xi,\eta)=\frac{1}{2}\,U\left(\xi^\top\eta+\eta^\top\xi\right)
+(1-\lambda)\Pi\left(\xi\eta^\top+\eta\xi^\top\right)U.
\label{Cf}
\end{equation}

Now, we present the solution of Equation~(\ref{geo1}) as a series
in $t$ in terms of the initial data $U(0)=U_0$ and ${\dot U}(0)=V$,
which are related by $U_0^\top V+V^\top U_0=0$. One has
\begin{equation}\label{UU}
U(t)=U_0+tV-\frac{t^2}{2!}\,\Gamma_0(V,V)-\frac{t^3}{3!}\,\Gamma_0(V,V,V)-\ldots,
\end{equation}
where $\Gamma_0(V,V)=\left.\Gamma(V,V)\right|_{U=U_0}$, and~$\Gamma_0(V,V,V)=\left.{\rm D}_V\Gamma(V,V)-2\Gamma(V,\Gamma(V,V))\right|_{U=U_0}$
 involves the derivative ${\rm D}_V$ at the point $U\in{\cal M}$ in the direction
$V\in{\cal TM}$, which acts, in~particular, as~%
\begin{eqnarray}
{\rm D}_\phi\Gamma(\xi,\eta)&=&\frac{1}{2}\,\phi\left(\xi^\top\eta+\eta^\top\xi\right)
-(1-\lambda)\left(\phi U^\top+U\phi^\top\right)\left(\xi\eta^\top+\eta\xi^\top\right)U
\nonumber\\
&&+(1-\lambda)\Pi\left(\xi\eta^\top+\eta\xi^\top\right)\phi.
\end{eqnarray}
Note that the expansions (\ref{xi}) and (\ref{DU}) can be applied {\it after}
calculating such a~derivative.

It is easy to verify in the approximation $t^2$ that $I_k=U^\top U=U_0^\top U_0$.
Thus, setting $t=1$, the~two points $U,U_0\in{\cal M}$ are related by (\ref{UU}).
Restoring the dependence of $U$ and $V$ on the coordinates $x$, formula (\ref{UU})
generalizes the ``field shift'' $U=U_0+V$ in Euclidean
space when $U,U_0,V\in\mathbb{R}^{\dim{\cal M}}$. Thus, fixing $U_0$,
the variables $V$ are called {\it normal coordinates}, which further describe
quantum fluctuations in the $\sigma$-model.

{\bf Curvature.} We define the curvature tensor $R(\xi,\eta)\phi$ in
$(1,3)$-form for $\xi,\eta,\phi\in{\cal TM}$ as (see~\cite{Ng22})
\begin{equation}\label{RT}
R(\xi,\eta)\phi={\rm D}_\eta\Gamma(\xi,\phi)-{\rm D}_\xi\Gamma(\eta,\phi)
+\Gamma(\eta,\Gamma(\xi,\phi))-\Gamma(\xi,\Gamma(\eta,\phi)).
\end{equation}

Let us calculate the components of $R(\xi,\eta)\xi=U(U^\top R(\xi,\eta)\xi)+\Pi R(\xi,\eta)\xi$:

~\vspace{6pt}
\begin{eqnarray}
U^\top R(\xi,\eta)\xi&=&-\frac{1}{4}\,(\omega_\eta \omega^2_\xi
-2\omega_\xi \omega_\eta \omega_\xi+\omega^2_\xi \omega_\eta)
+\frac{\lambda}{2}\,(\omega_\eta F^\top_\xi F_\xi+F^\top_\xi F_\xi \omega_\eta)
\nonumber\\
&&
+\frac{3-4\lambda}{4}\, (\omega_\xi F^\top_\xi F_\eta+F^\top_\eta F_\xi \omega_\xi)
-\frac{3-2\lambda}{4}\,(\omega_\xi F^\top_\eta F_\xi+F^\top_\xi F_\eta \omega_\xi),
\label{Rw}\\
\Pi R(\xi,\eta)\xi&=&-\lambda\,\frac{3-4\lambda}{2}\,F_\xi \omega_\xi \omega_\eta
+\lambda\,\frac{3-2\lambda}{2}\,F_\xi \omega_\eta \omega_\xi
-\lambda^2 F_\eta \omega^2_\xi+F_\eta F^\top_\xi F_\xi
\nonumber\\
&&
+\frac{2-3\lambda}{2}\,F_\xi F^\top_\xi F_\eta
-\frac{4-3\lambda}{2}\,F_\xi F^\top_\eta F_\xi.
\label{RF}
\end{eqnarray}

The bi-quadratic form $\widetilde{K}(\xi,\eta)=\langle R(\xi,\eta)\xi,\eta\rangle_{\rm g}$,
which characterizes the sectional curvature~\cite{KN}, can be expressed, after~substituting 
expressions (\ref{Rw}) and (\ref{RF}), as~follows:
\begin{eqnarray}
\widetilde{K}(\xi,\eta)&=&\frac{\lambda}{2}\,\left\|[\omega_\xi,\omega_\eta]\right\|^2_F
+\left\|F_\eta F^\top_\xi-F_\xi F^\top_\eta\right\|^2_F
+\frac{2-3\lambda}{2}\,\left\|F^\top_\eta F_\xi-F^\top_\xi F_\eta\right\|^2_F
\nonumber\\
&&
+2\lambda^2\left(\left\|F_\eta \omega_\xi\right\|^2_F+\left\|F_\xi \omega_\eta\right\|^2_F\right)
+2\lambda(3-4\lambda)\langle F_\xi \omega_\xi,F_\eta \omega_\eta\rangle_F
\nonumber\\
&&
-2\lambda(3-2\lambda)\langle F_\xi \omega_\eta,F_\eta \omega_\xi\rangle_F.
\end{eqnarray}
A similar formula is derived in the work (\cite{Jen75}, p.~405) in terms of
rectangular $(N-k)\times k$-matrices, say $M$, instead of $F\in\mathbb{R}^{N\times k}$.
However, it does not matter, because~$M$ and $F$ are entered as combinations
$M^\top M$ and $F^\top F$, which both belong to $\mathbb{R}^{k\times k}$.

Using (\ref{Rw})--(\ref{RF}) and contracting the $(1,3)$-curvature (see Appendix~\ref{AppA1}),
the diagonal components of the Ricci $(0,2)$-tensor are obtained~\cite{Ng22}:
\begin{equation}\label{RicL}
{\rm Ric}(\xi,\xi)=\left[\lambda^2(N-k)+\frac{k-2}{4}\right]\,
\Tr(\omega^\top_\xi\omega_\xi) +\left[N-2-\lambda(k-1)\right]\,
\Tr(F^\top_\xi F_\xi).
\end{equation}
It is seen that the curvature coefficients are constant due to the homogeneity
of the Stiefel manifold (see~\cite{Gav91}). We are discussing the relation
between $\widetilde{K}$ and ${\rm Ric}$ when renormalizing the $\sigma$-model.

\section{Background field~formalism}

We assume that the Lagrangian
${\cal L}[U_0]=\Tr({\bg\nabla}U_0^\top{\rm g}_0{\bg\nabla}U_0)$ is determined by
the bare metric ${\rm g}={\rm g}(U;\lambda)$, which differs from 
${\rm g}_0={\rm g}(U_0;\lambda)$ by the covariant counterterms ${\rm h}$:
\begin{equation}
{\rm g}=\mu^\varepsilon{\rm g}_0+{\rm h},
\end{equation}
where $\varepsilon=d-2$ is the renormalization parameter;
$\mu$ is  the renormalization~scale.

This renormalization constitutes our primary concern, though~the fields themselves
are also subject to renormalization. To~address this, we adapt a well-known procedure
(see~\cite{Muk81,How88}) to the model defined on the Stiefel manifold. Initially,
we present a covariant description utilizing normal~coordinates.

{\bf Effective Lagrangian.} Let us introduce the interpolating field $\Phi(x,s)$ and
quantities similar to those defined earlier:
\begin{equation}
\Phi(x,0)=U_0(x),\qquad
\left.\frac{\partial\Phi(x,s)}{\partial s}\right|_{s=0}=V,
\qquad
\Phi(x,1)=U(x).
\end{equation}

Let us expand the Lagrangian ${\cal L}[U]$ in a series with respect to
the parameter $s$:
\begin{equation}\label{Lexp}
{\cal L}[U]\equiv{\cal L}[\Phi(x,1)]=\sum\limits_{n=0}^\infty
\frac{1}{n!}\,\left.\frac{\partial^n{\cal L}[\Phi(x,s)]}{\partial s^n}\right|_{s=0},
\end{equation}
using the covariant derivative $\widehat{\nabla}_s$ along
the geodesic $\Phi(x,s)$, as~well as $\widehat{\bg\nabla}$:

~\vspace{6pt}
\begin{equation}
\widehat{\nabla}_s\xi\equiv{\frac{\partial\xi}{\partial s}}
+\Gamma\left(\frac{\partial\Phi(x,s)}{\partial s},\xi\right),
\qquad
\widehat{\bg\nabla}\xi\equiv{\bg\nabla}\xi
+\Gamma\left({\bg\nabla}\Phi(x,s),\xi\right),
\end{equation}
where the Christoffel function $\Gamma(\xi,\eta)$ from (\ref{Cf})
is taken at $U=\Phi(x,s)$.

Applying the decomposition of tangent vectors into $\omega$- and $F$-components,
we obtain
\begin{eqnarray}
\frac{\partial{\cal L}[\Phi(x,s)]}{\partial s}&=&
2\Tr\left({\bg\nabla}\Phi^\top{\rm g}\,\widehat{\bg\nabla}
\frac{\partial\Phi}{\partial s}\right),
\\
\frac{1}{2}\frac{\partial^2{\cal L}[\Phi(x,s)]}{\partial s^2}&=&
\Tr\left[\left(\widehat{\bg\nabla}\frac{\partial\Phi}{\partial s}\right)^\top
{\rm g}\,\widehat{\bg\nabla}\frac{\partial\Phi}{\partial s}\right]
+\Tr\left({\bg\nabla}\Phi^\top{\rm g}
\left[\widehat{\nabla}_s,\widehat{\bg\nabla}\right]\frac{\partial\Phi}{\partial s}\right),
\end{eqnarray}
taking into account that
\begin{eqnarray}
\widehat{\nabla}_s\frac{\partial\Phi(x,s)}{\partial s}&=&0,
\label{Rl1}\\
\widehat{\nabla}_s{\bg\nabla}\Phi(x,s)&=&\widehat{\bg\nabla}\frac{\partial\Phi(x,s)}{\partial s}
\equiv{\bg\nabla}\frac{\partial\Phi(x,s)}{\partial s}
+\Gamma\left(\frac{\partial\Phi(x,s)}{\partial s},{\bg\nabla}\Phi(x,s)\right),
\label{Rl2}\\
\left[\widehat{\nabla}_s,\widehat{\bg\nabla}\right]\xi
&=&R\left({\bg\nabla}\Phi(x,s),\frac{\partial\Phi(x,s)}{\partial s}\right)\xi,
\label{Rl3}
\end{eqnarray}
where the curvature $R(\xi,\eta)\phi$ is given by (\ref{RT}).

In the one-loop approximation, we restrict ourselves to the Lagrangian
quadratic in $V$:
\begin{equation}\label{L1}
{\cal L}[U]={\cal L}[U_0]
+2\langle{\bg\nabla}U_0,\widehat{\bg\nabla}V\rangle_{{\rm g}_0}
+\langle\widehat{\bg\nabla}V,\widehat{\bg\nabla}V\rangle_{{\rm g}_0}
+\langle{\bg\nabla}U_0, R({\bg\nabla}U_0,V)V\rangle_{{\rm g}_0},
\end{equation}
where $\widehat{\bg\nabla}V={\bg\nabla}V +\Gamma_0(V,{\bg\nabla}U_0)$,
and $\langle{\bg\nabla}U_0, R({\bg\nabla}U_0,V)V\rangle_{{\rm g}_0}=
-\widetilde{K}({\bg\nabla}U_0,V)$ by definition.
If $U_0(x)$ is a solution of the equation of motion, the~term
$\langle{\bg\nabla}U_0,\widehat{\bg\nabla}V\rangle_{{\rm g}_0}$
does not contribute to the action integral. Since expression (\ref{L1})
is general for $\sigma$-models within this approximation, we proceed
to the next step, which is standard for such~models.

We replace $V$ by ${\cal V}$ to transform
$\Tr({\bg\nabla}V^\top{\rm g}_0{\bg\nabla}V)$ into
$\Tr({\bg\nabla}{\cal V}^\top{\bg\nabla}{\cal V})$. To~do this,
we set $V=\widetilde{e}{\cal V}$, where $\widetilde{e}={\rm g}(U_0;\lambda^{-1/2})$.
Then $\widetilde{e}^2={\rm g}^{-1}_0$ and $\widetilde{e}^\top{\rm g}_0\widetilde{e}=I_N$,
and the relation between the components of the tangent vectors $V$ and ${\cal V}$
is given by
\begin{equation}
V=U_0\omega_V+F_V,\qquad
{\cal V}=U_0\omega_{\cal V}+F_V,\qquad
\omega_{\cal V}=\lambda^{1/2}\omega_V.
\end{equation}
After introducing $\omega_{\cal V}$ instead of $\omega_V$, the~quadratic part
of the Lagrangian with respect to $V$ takes on the form
\begin{eqnarray}
&{\cal L}_{\rm eff}=\langle\widehat{\bg\nabla}V,\widehat{\bg\nabla}V\rangle_{{\rm g}_0}
-\widetilde{K}({\bg\nabla}U_0,\widetilde{e}{\cal V}),&\\
&\langle\widehat{\bg\nabla}V,\widehat{\bg\nabla}V\rangle_{{\rm g}_0}=
\Tr\left({\bg\nabla}\omega^\top_{\cal V}{\bg\nabla}\omega_{\cal V}
+{\bg\nabla}F_V^\top{\bg\nabla}F_V\right)
+{\cal L}_{\rm int},&
\label{LV0}
\end{eqnarray}
where the first term in (\ref{LV0}) corresponds to the free evolution
of fluctuations and determines the Green's functions.
{In this free Lagrangian, we retain only the $F_V^Q$ components
of $F_V\in\mathbb{R}^{N\times k}$ as independent, since the $F^P_V$
components merely represent the interaction of $F_V^Q$ with $U_0$
(see Appendix \ref{AppC}).}

{\bf Quantization.} In fact, quantization here means
calculating the partition function and average values using
the methods of quantum theory, when the imaginary Planck constant
$-\rmi\hbar$ is replaced by the real temperature $T$. 
We consider $U_0(x)$
as the background field, and~$\omega_{\cal V}$ and $F_V$ as fluctuating
quantities. Simultaneously, the~components of the metric ${\rm g}$,
including the positive parameters $T$ and $\lambda$ (see (\ref{Lag})),
play the role of coupling constants. In~physical applications of
$\sigma$-models for describing Anderson localization, the~temperature 
$T$ is often associated with the conductivity coefficient, which
undergoes renormalization. On~the other hand, the~parameter $\lambda$
(anisotropy) appears in the Lagrangian (\ref{LwF}) as a running
(scale-dependent) factor.

Quantizing, one should introduce a path integral over the rapidly
varying fields $\omega_{\cal V}$ and $F_V$ (with a source), which is calculated
as a series 
 {using} Feynman diagrams. In~the case under consideration, all
expressions correspond to the one-loop approximation. As~usual, when averaging
expressions, Wick's theorem for pairing fields of the same sort does work and
allows us to limit ourselves to considering only even terms. In~principle,
the integral over the slowly varying component $U_0(x)$ also needs to be
applied, since it is ambiguous.
{Some computational aspects of this procedure
are outlined in Appendix \ref{AppC}.}

Let us define the Green's functions of the fields:
\begin{equation}\label{Fpair}
\llangle\omega_{\cal V}^{P_1}(x)\omega_{\cal V}^{P_2}(y)\rrangle=\pi_{P_1P_2}\Delta(x-y),
\qquad
\llangle F_V^{Q_1}(x) F_V^{Q_2}(y)\rrangle=\delta_{Q_1Q_2}\Delta(x-y),
\end{equation}
where the brackets $\llangle...\rrangle$ denote the quantum average in the vacuum state;
and the components of the matrix field $F_V$ with multi-indices $Q_n=(j_n,p_n)$, where
$1\leq p_n\leq k$ and $k+1\leq j_n\leq N$ are chosen as independent. In~addition,
the symbol $\delta_{Q_1 Q_2}=\delta_{j_1,j_2} \delta_{p_1,p_2}$, and~the symbol
\begin{equation}
\pi_{P_1P_2}=\frac{1}{2}
\left(\delta_{p_1,p_2}\delta_{q_1,q_2}-\delta_{p_1,q_2}\delta_{q_1,p_2}\right)
\end{equation}
for multi-indices $P_n=(p_n,q_n)$, $1\leq p_n,q_n\leq k$, and~their contraction
gives the number of degrees of freedom. The~spatially dependent component is
defined by the expression
\begin{equation}\label{FG}
\Delta(x-y)=T\int\frac{\rme^{\rmi k\cdot(x-y)}}{k^2}\,\frac{\rmd^d k}{(2\pi)^d},
\end{equation}
which results from (\ref{LV0}) and (\ref{dsq}).

\section{Beta functions of effective~couplings}

{\bf Renormalization.} Due to the form (\ref{FG}), when averaging local
expressions of the model, the~infrared (IR) divergences arise, which can
be mitigated by introducing a regulator.
{Softening the expressions at the IR boundary by using the scale $\mu$ and}
denoting the area of the unit sphere in $d$ dimensions as 
$\Omega_{d-1}=2\pi^{d/2}/\Gamma(d/2)$, when $\varepsilon=d-2\to 0$, we have
\begin{equation}\label{D0}
\Delta(0)=T\frac{\Omega_{d-1}}{(2\pi)^d}\int\limits_{\mu} k^{d-3}\rmd k
=-\frac{T}{2\pi\varepsilon}-\frac{T}{2\pi}\ln{\mu}+O(\varepsilon).
\end{equation}
At once, ultraviolet (UV) divergences are eliminated in a standard way
for $\sigma$-models~\cite{Bra85}.

Thus, the~metric renormalization in the one-loop approximation
is induced by the term~\cite{Muk81,How88,Bra85}
\begin{equation}\label{KR}
\int\llangle
\widetilde{K}({\bg\nabla}U_0,\widetilde{e}{\cal V})
\rrangle
\rmd^d x
=\int{\rm Ric}({\bg\nabla}U_0,{\bg\nabla}U_0)\,\Delta(0)\rmd^d x.
\end{equation}
The idea of proving formula (\ref{KR})  is based on the equality
of the quantum average of $\widetilde{K}$ over the fields
$\omega_{\cal V}$ and $F_V$ and the average over the basis matrices,
when ${\bg\nabla}U_0=U_0{\bg\omega}_u+{\bf F}_u$ (see Appendix~\ref{AppA1}).
That is, due to the homogeneity of the Stiefel manifold,
the calculations can be shifted to the ``origin'' ${\cal O}$, allowing
the use of the basis (\ref{base}). Then, we perform the substitution
$\omega_{\cal V}^P\to X_P$, $F_V^Q\to\Pi X_Q$, while the fields
themselves are absorbed by $\Delta$. We also express ${\bg\nabla}U_0$
in the basis (\ref{base}) and sum over $Q$ and the independent indices
of $P$. When averaging $\widetilde{K}({\bg\nabla}U_0,\widetilde{e}{\cal V})$,
only the terms quadratic in $\omega_{\cal V}$ and $F_V$ survive,
and we obtain
\begin{equation}\label{RicL2}
{\rm Ric}({\bg\nabla}U_0,{\bg\nabla}U_0)=\left[\lambda^2(N-k)+\frac{k-2}{4}\right]\,
\Tr({\bg\omega}^\top_u{\bg\omega}_u) +\left[N-2-\lambda(k-1)\right]\,
\Tr({\bf F}^\top_u {\bf F}_u).
\end{equation}

The divergences of (\ref{KR}) at $\varepsilon\to 0$ are eliminated by
the scaling factors $Z_{\rm g}$'s for the metric components:
\begin{equation}
Z_{{\rm g}_\omega}\frac{\lambda}{T}\delta_{P_1P_2},
\qquad
Z_{{\rm g}_F}\frac{1}{T}\delta_{Q_1Q_2}.
\end{equation}
According to (\ref{L1}), (\ref{D0}) and (\ref{RicL2}), we obtain
\begin{equation}
Z_{{\rm g}_\omega}=1-\frac{\tau}{\varepsilon}\left[\lambda^2(N-k)+\frac{k-2}{4}\right],
\qquad
Z_{{\rm g}_F}=1-\frac{t}{\varepsilon}[N-2-\lambda(k-1)],
\end{equation}
where we have defined the parameters $t=T/(2\pi)$ and $\tau=T/(2\pi\lambda)$.

{\bf Beta functions.} Renormalizing the model metric as $z=\ln{\mu}$ changes,
we demand
\begin{equation}
\frac{\rmd}{\rmd z}\ln{\left(\mu^\varepsilon Z_{{\rm g}_F}t^{-1}\right)}=0,\qquad
\frac{\rmd}{\rmd z}\ln{\left(\mu^\varepsilon Z_{{\rm g}_\omega}\tau^{-1}\right)}=0.
\end{equation}
These two conditions can be satisfied by assuming that both $t$ and $\lambda$
(or $\tau$) depend on $z$. Otherwise, to~renormalize only the temperature $t$,
we would have to require that the Stiefel manifold be Einsteinian when we equate
\begin{equation}\label{EinP}
\lambda^2(N-k)+\frac{k-2}{4}=\lambda[N-2-\lambda(k-1)].
\end{equation}
This is the same as $P_{N,k}(\lambda)=0$ for
$P_{N,k}(\lambda)\equiv\lambda^2(N-1)-\lambda(N-2)+(k-2)/4$.

Defining the beta functions $\beta_t=\rmd t/\rmd z$ and
$\beta_\lambda=\rmd\lambda/\rmd z$ that need to be found, we arrive at
the set of exact equations:
\begin{eqnarray}
&&\beta_t\left(\frac{1}{t}-\frac{\partial\ln{Z_{{\rm g}_F}}}{\partial t}\right)
-\beta_\lambda\,\frac{\partial\ln{Z_{{\rm g}_F}}}{\partial\lambda}=\varepsilon,\\
&&\beta_t\left(\frac{1}{t}-\frac{\partial\ln{Z_{{\rm g}_\omega}}}{\partial t}\right)
-\beta_\lambda\left(\frac{1}{\lambda}+\frac{\partial\ln{Z_{{\rm g}_F}}}{\partial\lambda}\right)
=\varepsilon.
\end{eqnarray}
Determining $Z_{\rm g}$'s up to order $t$, they reduce in the one-loop approximation to
\begin{equation}
\frac{\beta_t}{t}=\varepsilon Z_{{\rm g}_F},\qquad
\frac{\beta_t}{t}-\frac{\beta_\lambda}{\lambda}=\varepsilon Z_{{\rm g}_\omega},
\end{equation}
where $\beta_t/t-\beta_\lambda/\lambda=\beta_\tau/\tau$ for $\tau=t/\lambda$.

Note that taking into account all the terms in these equations using
approximate $Z_{\rm g}$'s may change the picture of the renormalization
group (RG) dynamics, but~will not correct the parameters of the stable
fixed point (sink) described~below.

Thus, we arrive at
\begin{equation}\label{betas}
\beta_t=\varepsilon t-[N-2-\lambda(k-1)]t^2,
\qquad
\beta_\tau=\varepsilon\tau-\left[\lambda^2(N-k)+\frac{k-2}{4}\right]\tau^2.
\end{equation}
Replacing $\beta_\tau$ with tantamount $\beta_\lambda=tP_{N,k}(\lambda)$,
we rewrite it in an equivalent form:
\begin{eqnarray}
\beta_\lambda&=&t(N-1)(\lambda-\lambda_+)(\lambda-\lambda_-),
\label{bL}\\
\lambda_\pm&=&\frac{N-2}{2(N-1)}\left[1\pm\sqrt{1-\frac{(k-2)(N-1)}{(N-2)^2}}\right].
\end{eqnarray}

\begin{figure}[h!]
\begin{center}
\includegraphics[width=11cm,angle=0]{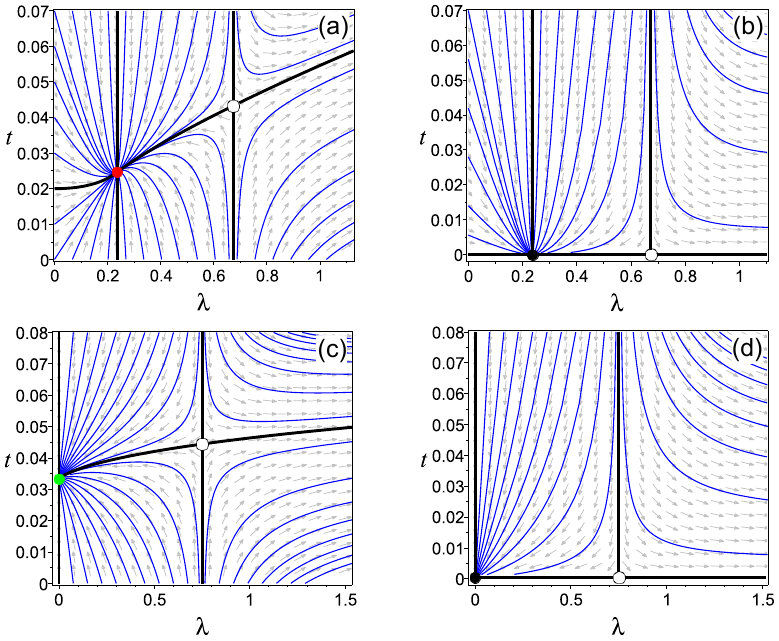}
\end{center}
\vspace*{-2mm}
\caption{\label{fig1}Phase regions of the model. {(\textbf{Top~row})} 
 $N=12$, $k=9$,
when $d=2.2$ in (\textbf{a}) and $d=2$ in (\textbf{b}). (\textbf{Bottom row}) $N=5$, $k=2$, when $d=2.1$
in (\textbf{c}) and $d=2$ in (\textbf{d}). Gray arrows denote velocity field. Colored circles are sinks,
while the white ones are saddle points. Black 
lines are phase separatrices, while blue curves are RG trajectories.}
\end{figure}

The region of admissible values of the model parameters in the plane $(\lambda;t)$
is the quadrant with $\lambda>0$ and $t>0$. This region is divided into three
subregions by two separatrices $(\lambda_-;t)$ and $(\lambda_+;t)$ for arbitrary $t>0$.
The fixed points are found from the conditions $\beta_t=0$ and $\beta_\lambda=0$.
Defining
\begin{equation}
t_\pm=\frac{\varepsilon}{N-2-\lambda_\pm(k-1)},
\end{equation}
we have the Gaussian point $(0;0)$, the~node $(\lambda_-;t_-)$,
and the saddle point $(\lambda_+;t_+)$ (see Figure~\ref{fig1}a).
\vspace{-3pt}

The trajectories of the renormalization group (RG) are given by the equations:
\begin{eqnarray}
\frac{\rmd t}{\rmd z}=\varepsilon t-[N-2-\lambda(k-1)]t^2,
&\quad &t(0)=t_0,
\label{eqt}\\
\frac{\rmd\lambda}{\rmd z}=t\left[\lambda^2(N-1)-\lambda(N-2)+\frac{k-2}{4}\right],
&\quad &\lambda(0)=\lambda_0.
\label{eql}
\end{eqnarray}
Then, the separatrices in Figure~\ref{fig1}, passing through the points $(0;\varepsilon⁄(N-2))$,
$(\lambda_-;t_-)$ and $(\lambda_+;t_+)$, are obtained numerically by integrating
(\ref{eqt}) and (\ref{eql}). It can be seen that in the case of $2<k<N$ and $d>2$,
the point $(\lambda_-;t_-)$, at~which the four phases meet, seems tetracritical
(red circle in Figure~\ref{fig1}a). Note that at $k=2$, such tetracriticality disappears,
and the point $(\lambda_-;t_-)=(0;\varepsilon⁄(N-2))$ becomes bicritical (green
circle in Figure~\ref{fig1}c). Finally, at~$d=2$ and fixed $\lambda$, phase
transitions by varying temperature $t$ are not observed
(see Figure~\ref{fig1}b,d), although~the sink--saddle pair still~exists.

\begin{figure}[h!]
\begin{center}
\includegraphics[width=5cm,angle=0]{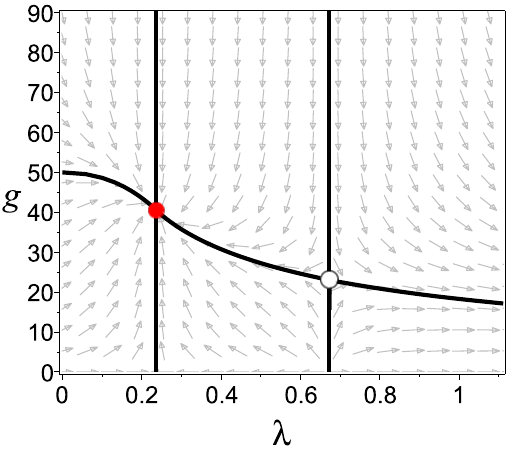}
\end{center}
\vspace*{-2mm}
\caption{\label{fig2}Phase regions of the model in the $(g,\lambda)$-plane
for $N=12$, $k=9$, $d=2.2$. The red and white circles, as before, indicate
the stable fixed point (sink) and the saddle point, respectively.}
\end{figure}

By introducing the quantity $g=1/t$, associated with the conductivity in some models,
and the corresponding beta function $\beta_g\equiv\rmd g/\rmd z=-g^2\beta_t(t=1/g)$,
we complement the case in Figure~\ref{fig1}a with the behavior of $g$ in Figure~\ref{fig2}.

{\bf Critical exponents.} As indicated above, the~critical point---a stable fixed
point (sink)---is given by the parameters $(\lambda_-;t_-)$. By~fixing $\lambda=\lambda_-$,
and thereby reducing the Stiefel manifold to an Einstein manifold (with proportionality
between the Ricci tensor and the metric), it is easy to calculate the critical exponent
$\nu$ (and $\eta$) in the one-loop approximation for $2\leq k<N$ and $\varepsilon>0$.
Under these conditions, there exists a positive critical temperature $t_c=t_-$,
the determination of which provides~\cite{Hi80}:
\begin{equation}
\frac{1}{\nu}=-\beta^\prime_t(t_c)=\varepsilon+O(\varepsilon^2).
\end{equation}
This agrees with the known result for the Grassmannian
$SO(N)⁄[SO(N-k)\times SO(k)]$ with the Einstein~property.

\vspace{-6pt}

Focusing primarily on the specifics of renormalizing the metric rather than the field,
based on the results in Appendix \ref{AppB} we may estimate the exponent
$\eta$ similarly to~\cite{BZJ76}:
\begin{eqnarray}
\eta&=&\zeta(t_c)-\varepsilon
\nonumber\\
&=&\varepsilon\left[\frac{N-k+(k-1)/(2\lambda_-)}{N-2-\lambda_-(k-1)}-1\right]
+O(\varepsilon^2).
\end{eqnarray}
We observe that $\eta\to0+O(\varepsilon^2)$ for large $N$, and~furthermore,
when $k=1$, this expression yields the previously established value~\cite{BZJ76}.

Other critical exponents can be found using the scaling and hyperscaling
relations:
\begin{equation}
\alpha=2-\nu d,\quad \beta=\nu\frac{d-2+\eta}{2},\quad
\gamma=(2-\eta)\nu,\quad \delta=\frac{d+2-\eta}{d-2+\eta}.
\end{equation}
Their values for the model in the one-loop approximation are given
in the Table~\ref{tab1}. It is worth noting that just the rightmost
column shows clearly the dependence on $k$: that is easily seen if
we put $N=12$ in the left neighboring (or $k=2$) column and then
compare the values in both columns. It is also interesting to note
that at $k=1$, the critical exponent $\beta$ does not depend on
$\varepsilon$, unlike all other exponents in this column. On~the other
hand, the~critical exponents $\alpha$ and $\nu$ do not depend on $N$
and are the same for any $k>0$.

\vspace{-3pt}
\begin{table}[h!] 
\caption{\label{tab1}{Critical} 
 exponents for the temperature transition at $\lambda=\lambda_-(N,k)$.}
\begin{center}
\begin{tabular}{|c|c|c|c|}
\hline
{Parameters} & {$N>2$, $k=1$} & {$N>2$, $k=2$} & {$N=12$, $k=9$}\\
\hline
{${d}$}
        & $2+\varepsilon$         & $2+\varepsilon$     & $2.2$\\
${t_c}$      & $\varepsilon/(N-2)$     & $\varepsilon/(N-2)$ & $0.0247$\\
${\alpha}$	   & $1-2/\varepsilon$       & $1-2/\varepsilon$   & $-9$ \\
${\beta}$     & $(N-1)/[2(N-2)]$        & $+\infty$           & $1.2278$ \\
${\gamma}$    & $2/\varepsilon-1/(N-2)$ & $-\infty$           & $8.5443$ \\
${\delta}$    & $\frac{(4+\varepsilon)N
-8-3\varepsilon}{\varepsilon(N-1)}$        & $-1$                & $7.9589$ \\
${\nu}$       & $1/\varepsilon$         & $1/\varepsilon$     & $5$ \\
${\eta}$      & $\varepsilon/(N-2)$     & $+\infty$           & $0.2911$ \\
\hline
\end{tabular}
\end{center}
\end{table}
\vspace{-3pt}

\section{Conclusions}

In this work, starting from the field Lagrangian for the $(2+\varepsilon)$-dimensional
nonlinear sigma model on the real Stiefel manifold, we applied the background-field method
and normal coordinates for the quantum treatment and obtained a renormalization group
description of the renormalized model in terms of two effective or running (depending
on the scaling parameter) charges. This result fully confirms the corresponding formulas for beta
functions and the properties of the renormalization group behavior of the studied model,
previously obtained~\cite{Gav89,Gav91} by applying a pure geometric approach to the general
case of the matrix beta function of a nonlinear model on an arbitrary Riemannian manifold.
The special tetracritical behavior---the presence of a tetracritical point common
to four phases---requires further detailed study (of the symmetry properties, etc.)
of each of these phases and the corresponding critical exponents. 
No less interesting is the subsequent application of the considered model to the
description of the properties of real tetracritical physical systems,
in particular to the problem of the possible coexistence and mutual influence of
the phenomenon of (de)localization in a quasi-two-dimensional electronic system and
the phenomenon of superconductivity, as~well as other systems with two independent
order parameters (see, e.g.,~Refs.~\cite{Birman,Sannino} for two such examples of
symmetry-based systems  that manifest tetracritical behavior). More generally speaking,
the obtained results may also be useful from the viewpoint of studying new classes 
of universality  and their systematics in the physics of (multi)critical~phenomena.

\vspace*{2mm} 

{\bf Acknowledgments.} Both authors acknowledge support from the National Academy of
Sciences of Ukraine by its priority project No.~0122U000888 and support from
the Simons Foundation grant 1290587.

\appendix
\section{Calculating the curvature tensors}

\subsection{\label{AppA1}Ricci curvature tensor}

Let us calculate the diagonal components of the Ricci $(0,2)$-tensor in
the basis (\ref{base}), using the contraction:
\begin{eqnarray}
{\rm Ric}(\xi,\xi)&=&\sum\limits_{\eta} \Tr\left[\eta^\top R(\xi,\eta)\xi\right]
\nonumber\\
&=&\sum\limits_{\omega_\eta, F_\eta} \Tr\left[\omega_\eta^\top U^\top R(\xi,\eta)\xi
+F^\top_\eta \Pi R(\xi,\eta)\xi\right].
\label{RicL1}
\end{eqnarray}

Due to the homogeneity of Stiefel manifolds, we perform calculations at
the ``origin'' ${\cal O}$, that is, with the matrix $U_{\cal O}$ and
the corresponding projector $\Pi_{\cal O}$:
\begin{equation}
U_{\cal O}=\left(\begin{array}{c} I_k \\ 0^{(N-k)\times k}\end{array}\right),
\qquad
\Pi_{\cal O}=I_N-U_{\cal O} U^\top_{\cal O}.
\end{equation}

Define the components of the tangent vector $\eta$ in the basis (\ref{base})
and at the ``origin'' as
\begin{equation}
\omega_\eta(P)=2^{-1/2}\,X_P,\qquad
F_\eta(Q)=\Pi_{\cal O} X_Q,
\end{equation}
where multi-indices $P=(p,q)$ and $Q=(j,r)$ consist of
$p,q,r=\overline{1,k}$ and $j=\overline{k+1,N}$. This means
that 1) we embed the components $\omega_\eta$ and $F_\eta$
in $\mathbb{R}^{N\times N}$, setting all remaining components
to zero; 2) we normalize $\eta$ such that
$\Tr[\omega^\top_\eta(P)\omega_\eta(P)]=1$ and
$\Tr[F^\top_\xi(Q) F_\xi(Q)]=1$.

Arbitrarily fixing the multi-indices $P_0$ and $Q_0$ of
the components $\omega_\xi$ and $F_\xi$, we write
\begin{equation}
\omega_\xi(P_0)=2^{-1/2}\,\omega\,X_{P_0},\qquad
F_\xi(Q_0)=F\Pi_{\cal O} X_{Q_0},
\end{equation}
where $\omega$ and $F$ are the values at ${\cal O}$ such that
$\Tr[\omega^\top_\xi(P_0)\omega_\xi(P_0)+F^\top_\xi(Q_0)
F_\xi(Q_0)]=\omega^2+F^2$.

Substituting the defined vector componets into (\ref{RicL1})
and numerically summing over $P=(p,q)$, $p<q$, and $Q=(j,r)$
in the corresponding terms, one gets
\begin{equation}
r(P_0,Q_0)=\left[\lambda^2(N-k)+\frac{k-2}{4}\right] \omega^2
+\left[N-2-\lambda(k-1)\right] F^2.
\end{equation}

Summing all terms $r(P_0,Q_0)$ with admissible $P_0$ and $Q_0$,
we obtain an expression for the diagonal components of the Ricci tensor:
\begin{equation}\label{RicL}
{\rm Ric}(\xi,\xi)=\left[\lambda^2(N-k)+\frac{k-2}{4}\right]\,
\Tr(\omega^\top_\xi\omega_\xi) +\left[N-2-\lambda(k-1)\right]\,
\Tr(F^\top_\xi F_\xi).
\end{equation}
Due to the homogeneity of Stiefel manifolds, the curvature coefficients
are constant at all points in space.

Decompose (\ref{RicL}) into orthogonal components for admissible
$P$ and $Q$ to obtain
\begin{equation}\label{RicLL}
{\rm Ric}(X_P,X_P)=2\lambda^2(N-k)+\frac{k-2}{2},
\qquad
{\rm Ric}(X_Q,X_Q)=N-2-\lambda(k-1),
\end{equation}
where we replace $\Tr(\omega^\top_\xi\omega_\xi)\to\Tr(X^\top_P X_P)=2$
and $\Tr(F^\top_\xi F_\xi)\to\Tr(X^\top_Q\Pi_{\cal O} X_Q)=1$.

We would also like to note another derivation of the tensors for $\lambda=1/2$
in Appendix~\ref{AppA2}.

\subsection{\label{AppA2}Curvature in the canonical case $\lambda=1/2$}

Focusing on the historically earlier and instructive case $\lambda=1/2$
for $X,Y,Z\in\frak{m}$, when $\frak{so}(N)\simeq\frak{m}\oplus\frak{h}$
at $\frak{h}=\frak{so}(N-k)$, let us analyze the curvature $R$ at the
``origin'' ${\cal O}$ (see, Proposition~3.4 in \cite{KN}, II, p.~202):
\begin{eqnarray}
(R(X,Y)Z)_{\cal O}&=&\frac{1}{4}\,[X,[Y,Z]_\frak{m}]_\frak{m}
-\frac{1}{4}\,[Y,[X,Z]_\frak{m}]_\frak{m}
\nonumber\\
&&-\frac{1}{2}\,[[X,Y]_\frak{m},Z]_\frak{m}
-[[X,Y]_\frak{h},Z],
\label{R}
\end{eqnarray}
where the commutator markings denote projections onto $\frak{m}$
and $\frak{h}$, and there is an obvious decomposition
$[X,Y]=[X,Y]_\frak{m}+[X,Y]_\frak{h}$.

Due to Proposition~3.4 from (\cite{KN}, II, p.~202), define the bi-quadratic form:
\begin{equation}
\widetilde{K}(X,Y)\equiv g(R(X,Y)Y,X)=\frac{1}{4}\,g([X,Y]_{\frak{m}},[X,Y]_{\frak{m}})
+g([X,Y]_{\frak{h}},[X,Y]_{\frak{h}}),
\end{equation}
which has the properties: $\widetilde{K}(Y,X)=\widetilde{K}(X,Y)$,
$\widetilde{K}(X,X)=0$, $\widetilde{K}(X+Y,Y)=\widetilde{K}(X,Y)$,
$\widetilde{K}(\alpha X,Y)=\widetilde{K}(X,\alpha Y)=\alpha^2 \widetilde{K}(X,Y)$,
where $\alpha$ is a scalar.

Taking into account the commutation relation (\ref{son}) and the fact that given $f^C_{AB}$
is not anti-symmetrized over indices $(c_1,c_2)=C$, we can write that $[X_A,X_B]_{\frak{m}}=
f^{(p,q)}_{AB}X_{(p,q)}+f^{(p,j)}_{AB}X_{(p,j)}+f^{(j,p)}_{AB}X_{(j,p)}\in\frak{m}$
and $[X_A,X_B]_{\frak{h}}=f^{(j_1,j_2)}_{AB}X_{(j_1,j_2)}\in\frak{h}$. Then, we have for
matrices $X_A,X_B\in\frak{m}$ that
\begin{eqnarray}
\widetilde{K}(X_A,X_B)&=&\frac{1}{4}\sum\limits_{p,q=1}^k(1-\delta_{p,q})
\left(f^{(p,q)}_{AB}\right)^2+\frac{1}{4}\sum\limits_{p=1}^k\sum\limits_{j=k+1}^N
\left[\left(f^{(p,j)}_{AB}\right)^2+\left(f^{(j,p)}_{AB}\right)^2\right]
\nonumber\\
&&+\sum\limits_{j_1,j_2=k+1}^N(1-\delta_{j_1,j_2})\left(f^{(j_1,j_2)}_{AB}\right)^2.
\label{KK}
\end{eqnarray}

Using $\widetilde{K}(X,Y)$, the {\it sectional curvature} of $V_k(\mathbb{R}^N)$
is determined by
\begin{equation}\label{secc}
K(X,Y)=\frac{\widetilde{K}(X,Y)}{\|X\|^2\|Y\|^2-g(X,Y)^2}.
\end{equation}
where the divisor equals unit in the case of orthonormal metric
$g(X_A,X_B)=\delta_{AB}$, see (\ref{mmm}).

The Riemann curvature tensor for $V_k(\mathbb{R}^N)$ is given by (\cite{KN}, I, p.~201):
\begin{equation}
R(X_1,X_2,X_3,X_4)=g(R(X_3,X_4)X_2,X_1).
\end{equation}
For instance, it can be presented for distinct $X_1,\dots,X_4$ as
\begin{eqnarray}
R(X_1,X_2,X_3,X_4)&=&\frac{1}{6}\left[\widetilde{K}(X_1+X_3,X_2+X_4)
-\widetilde{K}(X_2+X_3,X_1+X_4)\right.
\nonumber\\
&&\left.+\widetilde{K}(X_3,X_1)+\widetilde{K}(X_4,X_2)
-\widetilde{K}(X_3,X_2)-\widetilde{K}(X_4,X_1)\right].
\end{eqnarray}

Therefore, contracting the Riemann tensor, the Ricci curvature tensor
(\cite{KN}, I, p.~249) is determined by
\begin{equation}
R(X,Y)=\sum\limits_{Z\in\frak{m}}g(R(Z,X)Y,Z),
\end{equation}
Its diagonal components for $X\in\frak{m}$ are written as
\begin{eqnarray}
{\rm Ric}(X,X)&=&\sum\limits_{Z\in\frak{m}} \widetilde{K}(Z,X)\\
&=&\sum\limits_{p<q=1}^k\widetilde{K}(X_{(p,q)},X)
+\sum\limits_{p=1}^k\sum\limits_{j=k+1}^N\widetilde{K}(X_{(j,p)},X),
\end{eqnarray}
Using (\ref{KK}), one can numerically verify for arbitrary indices
$P=(p,q)$ and $Q=(j,p)$ that
\begin{equation}\label{RicC}
{\rm Ric}(X_P,X_P)=\frac{N-2}{2},\qquad {\rm Ric}(X_Q,X_Q)=N-2-\frac{k-1}{2}.
\end{equation}
These numerical coefficients coincide with (\ref{RicLL}) at $\lambda=1/2$.

\section{Certain aspects of quantization}

\subsection{\label{AppC}Path integral}

Using the action integral (\ref{Lag}), the partition function
${\cal Z}$ is given by the path integral with an invariant
measure defined up to a constant factor involving
$\sqrt{\det{{\rm g}(U(x);\lambda)}}=\lambda^{k/2}$:
\begin{equation}\label{PF1}
{\cal Z}=\int\exp{(-{\cal A}[U])}\,\delta(U^\top U-I_k)\,{\cal D}U,
\end{equation}
where the formal $\delta$-function fixes $k(k+1)/2$ constraints at each point $x$.

Note that, unlike quantum theory, where the imaginary Planck constant
$-\rmi\hbar$ is used, we employ low temperature $T$ as a measure of
positive definite action.

Introducing the on-shell background field $U_0(x)$, $U_0^\top U_0=I_k$,
which justifies the equation
${\bg\nabla}^2U_0+\Gamma_0({\bg\nabla}U_0,{\bg\nabla}U_0)=0$, we set
$U=U_0+\upsilon$, where the shift field $\upsilon$ is expressed in terms
of the normal coordinates $V\in{\cal T}{\cal M}$ according to (\ref{UU}):
\begin{equation}
\upsilon=V-\frac{1}{2!}\,\Gamma_0(V,V)-\frac{1}{3!}\,\Gamma_0(V,V,V)-\ldots.
\end{equation}

The orthogonality condition $U^\top U=I_k$ in terms of new variables is
satisfied due to the vanishing of $s_V\equiv U_0^\top V+V^\top U_0$. Then, 
decomposing $V=U_0(\omega_V+s_V)+F_V$, where $\dim{\omega_V}=k(k-1)/2$
and $\dim{F_V}=(N-k)k$, the $k(k+1)/2$ degrees of freedom of $s_V$
are eliminated by the $\delta$-functions from (\ref{PF1}). Thus, we are
left with $V=U_0\omega_V+F_V=\lambda^{-1/2}U_0\omega_{\cal V}+F_V$,
as it must be for a tangent vector.

However, we still need to parameterize $F_V\in\mathbb{R}^{N\times k}$,
$U_0^\top F_V=0$, by independent variables. By choosing for this
$\|F_V^Q\|\in\mathbb{R}^{(N-k)\times k}$, where the multi-index $Q=(j,p)$
with $1\leq p,q\leq k$ and $k+1\leq j\leq N$, we can express
the dependent components $F_V^P$ with $P=(p,q)$ as
\begin{equation}
F_V^{(p,q)}=\sum\limits_{j=k+1}^N E_0^{(p,j)}F_V^{(j,q)},\qquad
E_0^\top\equiv\left\|E_0^{(p,j)}\right\|=-\left\|U_0^{(p,q)}\right\|^{-1} \left\|U_0^{(j,q)}\right\|^\top,
\end{equation}
where one would expect $\|U_0^{(p,q)}\|=\|\pm(I_k-\widetilde{U}_0^\top\widetilde{U}_0)^{1/2}\|$
with $\widetilde{U}_0=\|U_0^{(j,p)}\|$ to obtain $U_0^\top U_0=I_k$.
Obviously, $E_0=\widetilde{U}_0=0$, when we take $U_0=U_{\cal O}$ at the ``origin''.

Let us emphasize that below by $F_V$ we mean independent components $F_V^Q$,
for example in the integration measure, unless otherwise specified.

Under these conditions, we come to the integral over fluctuations:
\begin{equation}\label{Om1}
\Omega[U_0]=\int\exp{({\cal A}[U_0]-{\cal A}[U_0+\upsilon])}\,
{\rm Det}\left\|\left(
\frac{\delta\upsilon(x)}{\delta\omega^P_{\cal V}(y)},\frac{\delta\upsilon(x)}{\delta F^Q_V(y)}
\right)\right\|\,
{\cal D}\omega_{\cal V}\,{\cal D}F_V.
\end{equation}

Further, we are using a perturbative approach based on (vacuum) averaging:
\begin{eqnarray}
\llangle(\ldots)\rrangle&=&{\cal Z}_0^{-1}\int(\ldots)\exp{(-S_0[\omega_{\cal V},F_V])}{\cal D}\omega_{\cal V}\,{\cal D}F_V,\\
{\cal Z}_0&=&\int\exp{(-S_0[\omega_{\cal V},F_V])}{\cal D}\omega_{\cal V}\,{\cal D}F_V,\\
S_0[\omega_{\cal V},F_V]&\equiv&\frac{1}{2T}\int \Tr\left({\bg\nabla}\omega^\top_{\cal V}{\bg\nabla}\omega_{\cal V}
+{\bg\nabla}F_V^\top{\bg\nabla}F_V\right)\rmd^dx.
\end{eqnarray}

It leads to $\Omega[U_0]={\cal Z}_0\,\llangle\exp{(-{\cal A}_{\rm int}[U_0,\omega_{\cal V},F_V])}\rrangle$
with
\begin{equation}
{\cal A}_{\rm int}[U_0,\omega_{\cal V},F_V]\equiv{\cal A}[U_0+\upsilon]-{\cal A}[U_0]-S_0[\omega_{\cal V},F_V]
-\ln{D},
\end{equation}
where $D$ denotes the Jacobi determinant from (\ref{Om1}).

According to the Dyson--Wick algorithm, let us introduce the generating
functional with a skew-symmetric source $J$:
\begin{eqnarray}
{\cal Z}_0[J]&\equiv&{\cal Z}_0^{-1}\int
\exp{\left[-S_0+\int\Tr\left(J_\omega^\top\omega_{\cal V}
+J_F^\top F_V\right)\rmd^dx\right]}\,{\cal D}\omega_{\cal V}\,{\cal D}F_V
\nonumber\\
&=&\exp{\left[\frac{T}{2}\int\Tr\left(J_\omega^\top{\bg\nabla}^{-2}J_\omega
+J_F^\top{\bg\nabla}^{-2}J_F\right)\rmd^dx\right]}.
\end{eqnarray}
The Green's function (\ref{FG}) is resulted from
\begin{equation}\label{dsq}
{\bg\nabla}^{-2}\delta^d(x)=-\int\frac{\rme^{\rmi k\cdot x}}{k^2}\,\frac{\rmd^d k}{(2\pi)^d}.
\end{equation}

Thus, by expanding the interaction functional, the multi-point Green's function is
\begin{eqnarray}
&&\llangle\omega^{P_1}_{\cal V}(x_1)\ldots\omega^{P_n}_{\cal V}(x_n)F^{Q_1}_V(y_1)\ldots F^{Q_m}_V(y_m)\rrangle
\nonumber\\
&&=\left.\frac{\delta^{n+m}{\cal Z}_0[J]}{
\delta\left(J^{Q_m}_F(y_m)\right)^\top\ldots\delta\left(J^{Q_1}_F(y_1)\right)^\top
\delta\left(J^{P_n}_\omega(x_n)\right)^\top\ldots\delta \left(J^{P_1}_\omega(x_1)\right)^\top}
\right|_{J_\omega=0, J_F=0}.\qquad {\ }
\end{eqnarray}
Besides, the connected Green's functions are generated by the functional $-T\ln{\Omega[U_0]}$.

\subsection{\label{AppB}Renormalizing the field}

Renormalizing, due to introducing the field scaling factor $Z_U$, let us
require finiteness of the anomalous quantum mean $\llangle Z^{1/2}_U U\rrangle$
of the field $U\in V_k(\mathbb{R}^N)$ at the point $x\in\mathbb{R}^d$. Given
the background field $U_0\in V_k(\mathbb{R}^N)$, and using the normal
coordinates $V$ (see, for instance, (\ref{UU})), in the one-loop
approximation we can write that
\begin{eqnarray}
\llangle U\rrangle&=&\llangle U_0+V-\frac{1}{2!}\,\Gamma_0(V,V)\rrangle
\nonumber\\
&=&U_0\left[I_k-\frac{1}{2}\llangle\lambda^{-1}\omega^\top_{\cal V}\omega_{\cal V}
+F^\top_V F_V\rrangle\right],
\end{eqnarray}
where $\llangle U_0\rrangle=U_0$, and the Christoffel function, averaged as
$\llangle\Gamma_0(V,V)\rrangle=U_0\llangle V^\top V\rrangle$, is expressed in terms
of the $\omega$- and $F$-components of the field $V$ with $\llangle V\rrangle=0$.

According to the rules (\ref{Fpair}) and (\ref{D0}), we extract the divergent
part
\begin{equation}
\llangle U\rrangle\sim\left[1+\frac{t}{2\varepsilon}\left(\frac{k-1}{2\lambda}+N-k\right)\right]U_0,
\end{equation}
that immediately leads to
\begin{equation}
Z_U=1-\frac{t}{\varepsilon}\left(\frac{k-1}{2\lambda}+N-k\right).
\end{equation}
In particular, the substitution $k=1$ reproduces the result for the vector model \cite{BZJ76}.

Then, the scaling properties in our approximation can be given by the ``beta'' function:
\begin{equation}
\zeta=-\frac{\rmd\ln{Z_U}}{\rmd z}=\left(\frac{k-1}{2\lambda}+N-k\right)t.
\end{equation}


\end{document}